\documentclass[12pt]{iopart}
\usepackage{iopams}
\usepackage{graphicx}
\usepackage[dvips]{color}
\usepackage{amssymb, amsfonts, amsbsy}
\usepackage{mathrsfs} 
\usepackage{daytime}
\usepackage{dayofweek}
\begin{document}
\def\TODAY{13 June 2008}
\title[Bounding the Hubble flow in terms of the $w$ parameter]    
{Bounding the Hubble flow in terms of the $w$ parameter}
\author{
C\'eline Catto\"en and 
Matt Visser}
\address{School of Mathematics, Statistics, and Computer Science, \\
Victoria University of Wellington, PO Box 600, Wellington, New Zealand}
\ead{celine.cattoen@mcs.vuw.ac.nz, 
matt.visser@mcs.vuw.ac.nz}
\date{\TODAY; \LaTeX-ed \DayOfWeek, \today; \daytime}    
\begin{abstract}

The last decade has seen increasing efforts to circumscribe and bound the cosmological Hubble flow in terms of model-independent constraints on the cosmological fluid --- such as, for instance, the classical energy conditions of general relativity. Quite a bit can certainly be said in this regard, but much more refined bounds can be obtained by placing more precise constraints (either theoretical or observational) on the cosmological fluid. In particular, the use of the $w$-parameter ($w=p/\rho$) has become increasingly common as a \emph{surrogate} for trying to say something about the cosmological equation of state. Herein we explore the extent to which a constraint on the $w$-parameter leads to useful and nontrivial constraints on the Hubble flow, in terms of constraints on density $\rho(z)$, Hubble parameter $H(z)$, density parameter $\Omega(z)$, cosmological distances $d(z)$, and lookback time $T(z)$. In contrast to other partial results in the literature, we carry out the computations for arbitrary values of the space curvature $k\in[-1,0,+1]$, equivalently for arbitrary $\Omega_0 \lessgtr 1$.

\vskip 0.250cm

\noindent
Keywords:

\vskip 0.1250cm

\noindent arXiv 0806.nnnn [gr-qc];  \TODAY;  \\
\LaTeX-ed \DayOfWeek, \today; \daytime.

\end{abstract}
\maketitle
\newtheorem{theorem}{Theorem}
\newtheorem{corollary}{Corollary}
\newtheorem{lemma}{Lemma}
\def\d{{\mathrm{d}}}
\def\implies{\Rightarrow}
\newcommand{\scri}{\mathscr{I}}
\newcommand{\sun}{\ensuremath{\odot}}

\def\eg{{\it e.g.}}
\def\ie{{\it i.e.}}
\def\etc{{\it etc.}}
\def\sign{{\hbox{sign}}}
\def\eof{\Box}
\newenvironment{warning}{{\noindent\bf Warning: }}{\hfill $\eof$\break}
\tableofcontents
\markboth{Bounding the Hubble flow in terms of the $w$ parameter}{}
\clearpage

\markboth{Bounding the Hubble flow in terms of the $w$ parameter}{}

\section{Introduction}

The classical energy conditions of general relativity~\cite{Hell,Lorentzian}, despite their well-known limitations~\cite{Twilight, brane, quantifying}, are nevertheless very useful surrogates for controlling the extent to which one is willing to countenance extreme and unusual physics. As applied to cosmology, in addition to the very general cosmological singularity theorem presented in~\cite{Hell}, the classical energy conditions have (at the cost of additional hypotheses) been used to place more precise limits on the expansion of an idealized FLRW universe~\cite{science, prd, mg8, bounce, cosmo99}. More recently, these ideas have been extended in various ways in~\cite{more-on-ec, f-of-R, celine-rip, celine-ec}. In all of these analyses there is a trade-off between the \emph{precision} and \emph{generality} of the constraints one obtains --- the art lies in choosing a form of the input assumptions that is as general as possible, but not too general, for the precision of the output constraints one wishes to derive.

In the current article we shall derive some very general bounds in terms of assumptions about the $w$-parameter, where as usual $w=p/\rho$. Specifically, we shall ask the question: If we know for theoretical reasons, or can observationally determine, that $w$ lies in some restricted range
\begin{equation}
w(z) \in [w_-, w_+],
\end{equation}
between redshift zero and redshift $z$, 
what constraint does that place on the cosmological expansion? We shall see that considerable useful information can be extracted regarding the density $\rho(z)$, Hubble parameter $H(z)$, density parameter $\Omega(z)$, various cosmological distances $d_X(z)$, and lookback time $T(z)$.  Specifically, for some generic cosmological parameter $X(z)$,  we shall be looking for bounds of the form
\begin{equation}
X(z) \gtrless X_0 \; f(\Omega_0,z).
\end{equation}
Conversely, observational constraints on these cosmological parameters can be used to infer features of the cosmological fluid in a largely model-independent manner.  In contrast to other partial results scattered throughout the literature, we carry out the computations for arbitrary values of the space curvature $k\in[-1,0,+1]$, equivalently for arbitrary $\Omega_0 \lessgtr 1$.

\section{Strategy}

Our strategy will be to adopt a standard FLRW cosmology
\begin{equation}
\d s^2 = - c^2 \, \d t^2 
+ a(t)^2 \left\{ {\d r^2\over 1 - k r^2} + r^2 (\d\theta^2 + \sin^2\theta\,\d\phi^2) \right\},
\end{equation}
then,  (setting $8\pi G_N\to 1$, but explicitly retaining the speed of light $c$), we have the two Friedmann equations:
\begin{equation}
\rho = 3 \left[ {\dot a^2\over a^2} + {k c^2\over a^2} \right],
\qquad
\hbox{and}
\qquad
p  = - {\dot a^2\over a^2} - {k c^2\over a^2} - 2 {\ddot a\over a}.
\end{equation}
Together, these two equations imply the standard conservation law:
\begin{equation}
\dot \rho = -3(\rho+p) {\dot a\over a}.
\end{equation}
We also have the fundamental \emph{definitions}~\footnote{Historically it was common to refer to this quantity as the ``critical density'', $\rho_\mathrm{critical}$, but with the advent of widespread acceptance of a nonzero cosmological constant, or more generally ``dark energy'', the logical connection between this ``critical'' density and possible re-collapse of the universe has been severed. In a modern context then, it is inappropriate to refer to this as a ``critical'' density, and the considerably more neutral phrase ``Hubble density'' is preferable.} 
\begin{equation}
\rho_\mathrm{Hubble} = 3 \left[ {\dot a^2\over a^2} \right] = 3 H^2,
\end{equation}
and
\begin{equation}
\Omega = {\rho\over\rho_\mathrm{Hubble}} = {\rho\over3H^2} =  {H^2+ k c^2/a^2\over H^2} = 1 + {kc^2\over a^2 H^2}.
\end{equation}
For intermediate steps of the calculation we shall work with the very simple linear equation of state
\begin{equation}
p = w_* \; \rho,
\end{equation}
where $w_*$ is taken to be a constant. Picking some generic cosmological parameter $X(z)$,  we shall first calculate $X_{w_*}(z)$, and then (by assuming that $w(z)\in [w_-,w_+]$ from redshift zero out to redshift $z$, and depending on the direction of the relevant inequality) use this to derive bounds of the form
\begin{equation}
X_{w_-}(z)  \leq X(z) \leq X_{w_+}(z),
\end{equation}
or
\begin{equation}
X_{w_+}(z)  \leq X(z) \leq X_{w_-}(z).
\end{equation}
We shall also make the extremely mild assumption that the density is positive
\begin{equation}
\rho>0.
\end{equation}
This is certainly a completely redundant assumption for $k=0$ and $k= +1$ FLRW universes. Only for $k=-1$ universes does this provide the \emph{extremely mild} additional constraint $H>c/a$, that is, $H(z) > (c/a_0) \, (1+z)$.~\footnote{This is equivalent to enforcing $\dot a > c$, for a $k=-1$ FLRW universe, noting that $\dot a$ is \emph{not} a physical velocity, so that  $\dot a > c$ is a perfectly acceptable physical statement.}

\section{Density}

We now apply this strategy to the density. 
From
\begin{equation}
\dot \rho = -3(\rho+p) {\dot a\over a} = - 3 \rho ( 1+ w_*) {\dot a\over a},
\end{equation}
we have
\begin{equation}
{\dot \rho \over \rho} =  - 3 ( 1+ w_*) {\dot a\over a}.
\end{equation}
So integrating, for constant $w_*$ we obtain the well-known result
\begin{equation}
\rho_{w_*} =   \rho_0 (a/a_0)^{- 3 ( 1+ w_*)} =    \rho_0 \; (1+z)^{3 ( 1+ w_*)}.
\end{equation}
But now ask what happens if we only know that $w_- \leq w(z) \leq w_+$? (Where in the real observable universe $w(z)$ certainly need not be a constant.) Following the above analysis, we find that we must replace equalities by inequalities and so deduce 
\begin{equation}
\rho_0 \; (1+z)^{3 ( 1+ w_-)} \leq \rho(z) \leq   \rho_0 \; (1+z)^{3 ( 1+ w_+)}.
\qquad\qquad (z\geq 0).
\end{equation}
Note that for $z>0$ we are looking into the past; in contrast for $-1<z<0$, we are looking into the future~\cite{redshift}, and the inequality reverses to \footnote{Furthermore, note that there is no reason to ever go below $z=-1$, as $z=-1$ corresponds to infinite expansion. Also, note that the \emph{sign} of $1+w_-$ and $1+w_+$ does not affect these inequalities.}
\begin{equation}
\rho_0 \; (1+z)^{3 ( 1+ w_+)} \leq \rho(z) \leq   \rho_0 \; (1+z)^{3 ( 1+ w_-)};
\qquad\qquad (-1<z\leq 0).
\end{equation}
Of course, these simple constraints on the density are by far the most elementary of the inequalities we shall deduce --- some of the other inequalities derived below will prove to be  much more subtle.

If we now in addition relax our initial constraint on $\rho_0$, by assuming we only know that the present epoch density lies in some bounded interval
\begin{equation}
\rho_0 \in [\rho_{0_-}, \rho_{0_+}], \qquad \hbox{that is,} \qquad
\rho_{0_-} \leq \rho_0 \leq \rho_{0_+},
\end{equation}
then these two bounds generalize to
\begin{equation}
\rho_{0_-} (1+z)^{3 ( 1+ w_-)} \leq \rho(z) \leq   \rho_{0_+} (1+z)^{3 ( 1+ w_+)};
\qquad\quad (z\geq 0);
\end{equation}
\begin{equation}
\rho_{0_-} (1+z)^{3 ( 1+ w_+)} \leq \rho(z) \leq   \rho_{0_+} (1+z)^{3 ( 1+ w_-)};
\qquad\quad (-1<z\leq 0).
\end{equation}

\section{Density  parameter}

We have the following \emph{identity}
\begin{eqnarray}
\Omega -1 &\equiv& \frac{k\,c^2}{a^2\,H^2} =  \frac{k\,c^2}{a_0^2\,H_0^2} \; \frac{a_0^2}{a^2}\; \frac{H_0^2}{H^2} 
=\left( \Omega_0-1 \right) \; {\Omega\over\Omega_0} \; {\rho_0\over\rho}.
\end{eqnarray}
This leads to the useful result
\begin{equation}
{\Omega(z) - 1 \over\Omega(z)} = \left( {\Omega_0-1\over\Omega_0} \right) \; \frac{\rho_0}{\rho(z)}.
\end{equation}
Therefore, a bound on $\rho(z)$ automatically implies a bound on $\Omega(z)$.
From the result for $\rho_{w_*}(z)$ presented above, 
we deduce that bounds on $\Omega(z)$ can be given in terms of
\begin{equation}
{\Omega_{w_*}(z) - 1 \over\Omega_{w_*}(z)} = \left( {\Omega_0-1\over\Omega_0} \right) \; (1+z)^{-(3w_*+1)},
\label{E:omega}
\end{equation}
which we can equivalently recast as 
\begin{equation}
\Omega_{w_*}(z) = \frac{\Omega_0 \left( 1+z\right)^{3w_*+1} }{(1-\Omega_0) +\Omega_0  \left( 1+z\right)^{3w_*+1}}.
\end{equation}
We can now use this quantity, which was derived for strictly constant $w_*$,  to bound the density parameter $\Omega(z)$ for a more
realistic matter model satisfying the milder condition $w_- \leq w(z) \leq w_+$. We obtain:
\begin{itemize}

\item 
If $\Omega_0<1$  (but remember that by assumption $\Omega_0>0$) then
\begin{equation}
\Omega_{w_-}(z) \leq \Omega(z) \leq \Omega_{w_+}(z);
\qquad\qquad
(z>0),
\end{equation}
\begin{equation}
\Omega_{w_+}(z) \leq \Omega(z) \leq \Omega_{w_-}(z);
\qquad\qquad
(-1<z<0).
\end{equation}

\item 
If $\Omega_0=1$ then $\forall z : \Omega(z)=1$.

\item
If $\Omega_0>1$,
\begin{equation}
\Omega_{w_+}(z) \leq \Omega(z) \leq \Omega_{w_-}(z);
\qquad\qquad
(z>0),
\end{equation}
\begin{equation}
\Omega_{w_-}(z) \leq \Omega(z) \leq \Omega_{w_+}(z);
\qquad\qquad
(-1<z<0),
\end{equation}
but note that  the bound can break down when the denominator of $\Omega_{w_*}(z)$ equals zero --- this occurs at 
\begin{equation}
z_\Omega({w_*},\Omega_0) = \left({\Omega_0-1\over\Omega_0}\right)^{1/(3w_*+1)} - 1.
\end{equation}
The failure of the bound might occur either in the past or the future depending on the value of $w_*$. 
\begin{itemize}
\item If $3w_*+1> 0$ then the bound is useful only for  $z>z_\Omega(w_*,\Omega_0) < 0$, implying a limitation in the past.
\item If $3w_*+1=0$ then the bound is valid for all $z$.
\item If $3w_*+1< 0$ then the bound is useful only for  $z<z_\Omega({w_*},\Omega_0) > 0$, implying a limitation in the future. 
\end{itemize}
Note that nothing unusual need happen to the universe itself at $z_\Omega({w_*},\Omega_0)$, it is only the \emph{bound} that loses its predictive usefulness. Combining these observations we see that for $\Omega_0>1$ it is better (in the sense of reducing the amount of special case exceptions to the general rule) to recast the bounds in the form:
\begin{equation}
\fl
 \left( {\Omega_0-1\over\Omega_0} \right) \; (1+z)^{-(3w_++1)} \leq
 {\Omega(z) - 1 \over\Omega(z)} \leq
  \left( {\Omega_0-1\over\Omega_0} \right) \; (1+z)^{-(3w_-+1)} 
\end{equation}
for $z>0$, and 
\begin{equation}
\fl
 \left( {\Omega_0-1\over\Omega_0} \right) \; (1+z)^{-(3w_-+1)} \leq
 {\Omega(z) - 1 \over\Omega(z)} \leq
  \left( {\Omega_0-1\over\Omega_0} \right) \; (1+z)^{-(3w_++1)}
\end{equation}
for $-1<z<0$.
\end{itemize}

\section{Hubble parameter}

Let us now use the density equation (the first Friedmann equation) and the definition of the density parameter $\Omega$ to write
\begin{equation}
H^2 = {\rho\over3} - {kc^2\over a^2} =  {\rho\over\rho_0} {\rho_0\over3} -  {a_0^2\over a^2} {kc^2\over a_0^2} 
=  {\rho\over\rho_0} \Omega_0 H_0^2 -  {a_0^2\over a^2} (\Omega_0-1) H_0^2.
\end{equation}
That is, as an \emph{identity}:
\begin{equation}
H^2 = H_0^2 \left\{ \Omega_0  {\rho\over\rho_0} -  (\Omega_0-1) {a_0^2\over a^2} \right\} 
= H_0^2 \left\{ \Omega_0  {\rho\over\rho_0} -  (\Omega_0-1) (1+z)^2  \right\}.
\end{equation}
But we have already derived a formula for $\rho_{w_*}(z)$,
whence
\begin{equation}
H_{w_*}^2(z) =
 H_0^2 \left\{ \Omega_0 (1+z)^{3 ( 1+ w_*)} -  (\Omega_0-1) (1+z)^2  \right\},
\end{equation}
which we can recast as
\begin{equation}
H_{w_*}(z) =
 H_0 (1+z) \sqrt{ 1 -\Omega_0 + \Omega_0 (1+z)^{3w_*+1}  }.
\end{equation}
For realistic matter, satisfying some constraint $w_- \leq w(z) \leq w_+$, we then deduce
\begin{equation}
H_{w_-}(z) \leq H(z) \leq H_{w_+}(z);  \qquad (z>0);
\end{equation}
\begin{equation}
H_{w_+}(z) \leq H(z) \leq H_{w_-}(z);  \qquad (-1<z<0).
\end{equation}
Note that the Hubble bound ceases to provide useful information once the argument of the square root occurring in $H_{w_*}(z)$ becomes negative. 
\begin{itemize}
\item For $\Omega_0\leq 1$ there is no limitation in the physical region $z\in(-1,\infty)$.
\item
For $\Omega_0>1$ this limitation manifests itself at $z_H(w_*,\Omega_0)=z_\Omega(w_*,\Omega_0)$, the same place that the bound on $\Omega(z)$ ran into difficulties.
(Some numerical estimates of where the bounds fail, based on current consensus observational data, are discussed in~\cite{celine-ec}.)
\end{itemize}
Finally, suppose that we do not have precise information regarding $H_0$ and $\Omega_0$, and only have the more limited information
\begin{equation}
H_0 \in [H_{0_-}, H_{0_+}], \qquad \Omega_0 \in [\Omega_{0_-}, \Omega_{0_+}],
\end{equation}
then these two Hubble bounds further generalize to
\begin{eqnarray}
&&H_{0_-} (1+z) \sqrt{ 1 -\Omega_{0_-} + \Omega_{0_-} (1+z)^{3w_-+1}  }  
\leq H(z) 
\nonumber\\
&&
\qquad\qquad 
\leq
H_{0_+} (1+z) \sqrt{ 1 -\Omega_{0_+} + \Omega_{0_+} (1+z)^{3w_++1}  } ;
\quad
(z>0);
\end{eqnarray}
\begin{eqnarray}
&&H_{0_-} (1+z) \sqrt{ 1 -\Omega_{0_+} + \Omega_{0_+} (1+z)^{3w_++1}  }  
\leq H(z) 
\nonumber\\
&& 
\qquad
\leq
H_{0_+} (1+z) \sqrt{ 1 -\Omega_{0_-} + \Omega_{0_-} (1+z)^{3w_-+1}  } ;
\quad\quad
(-1<z<0);
\end{eqnarray}
subject to the caveat that for $\Omega_0>1$ we should not push the bound past $z_H(w_*,\Omega_0)$.

\section{Cosmological distances}

Let us for the time being focus on Peebles' definition of ``angular diameter distance''. This is what Weinberg calls the ``proper motion distance''~\cite{Peebles, Weinberg}, for more definitions and a discussion regarding the physical interpretation of the cosmological distance scales see~\cite{ Hogg}, see also~\cite{redshift, controversy}. We make this choice to minimize the number of factors of $1+z$ in subsequent formulae.  Then the standard definition is
\begin{equation}
d_P = a_0  \; \sin_k \left( {c\over a_0 H_0 }\int {H_0\over H(z)} \d z \right).
\end{equation}
But since
\begin{equation}
{c\over a_0 H_0} = \sqrt{k(\Omega_0-1)},
\end{equation}
this can be rewritten more suggestively as
\begin{equation}
d_P = {c\over H_0} {1\over \sqrt{1-\Omega_0}} 
\sinh  \left(    \sqrt{1-\Omega_0}\; \int {H_0\over H(z)} \d z \right).
\end{equation}
When interpreting this last formula for $\Omega_0>1$ we make use of the fact that $\sinh(i\Theta) = i \sin(\Theta)$. 
Substituting $H(z) \to H_{w_*}(z)$ and performing the integral, after considerable effort  both {\sf Mathematica} and  {\sf Maple} yield 
\begin{equation}
\fl
\int {H_0\over H_{w_*}(z)} \d z  = {2\over\sqrt{1-\Omega_0}\,(3w_*+1)}
\ln\left\{
{
(\sqrt{1-\Omega_0}+1) \; (1+z)^{(3w_*+1)/2}
\over
\sqrt{1-\Omega_0}+ \sqrt{1-\Omega_0+\Omega_0(1+z)^{(3w_*+1)}}
}
\right\},
\end{equation}
whence
\begin{eqnarray}
\fl
d_{P_{w_*}}(z) &=&  {c\over2H_0\sqrt{1-\Omega_0}} \left[
\left\{
{
(\sqrt{1-\Omega_0}+1) \; (1+z)^{(3w_*+1)/2}
\over
\sqrt{1-\Omega_0}+ \sqrt{1-\Omega_0+\Omega_0(1+z)^{(3w_*+1)}}
}
\right\}^{2/(3w_*+1)} \right.
\nonumber\\
\fl
&&
\qquad
-
\left.
\left\{
{
(\sqrt{1-\Omega_0}+1) \; (1+z)^{(3w_*+1)/2}
\over
\sqrt{1-\Omega_0}+ \sqrt{1-\Omega_0+\Omega_0(1+z)^{(3w_*+1)}}
}
\right\}^{-2/(3w_*+1)} \right].
\end{eqnarray}
This simplifies slightly
\begin{eqnarray}
\fl
d_{P_{w_*}}(z) &=& {c\over2H_0\sqrt{1-\Omega_0}} \left[
(1+z) \; \left\{
{
(\sqrt{1-\Omega_0}+1 ) 
\over
\sqrt{1-\Omega_0} + \sqrt{1-\Omega_0+\Omega_0(1+z)^{(3w_*+1)}}
}
\right\}^{2/(3 w_*+1)} \right.
\nonumber\\
\fl
&&
- \left.
(1+z)^{-1} \;\left\{
{
(\sqrt{1-\Omega_0}+1) 
\over
\sqrt{1-\Omega_0} + \sqrt{1-\Omega_0+\Omega_0(1+z)^{(3w_*+1)}}
}
\right\}^{-2/(3w_*+1)}
\right].
\end{eqnarray}
We now note
\begin{equation}
\fl
{
(\sqrt{1-\Omega_0}+1 ) 
\over
\sqrt{1-\Omega_0} + \sqrt{1-\Omega_0+\Omega_0(1+z)^{(3w_*+1)}}
}
= 
{
\sqrt{1-\Omega_0+\Omega_0(1+z)^{(3w_*+1)}} -  \sqrt{1-\Omega_0} 
\over
( 1 - \sqrt{1-\Omega_0} ) \; (1+z)^{3w_*+1}
},
\end{equation}
(cross multiply top and bottom), which finally permits us to write the most tractable form of our \emph{exact} result for Peebles' angular diameter distance (in a constant $w(z)=w_*$ FLRW universe): 
\begin{eqnarray}
\fl
d_{P_{w_*}}(z) &=& {c\over2H_0\sqrt{1-\Omega_0} \; (1+z) } \left[
\left\{
{
\sqrt{1-\Omega_0+\Omega_0(1+z)^{(3w_*+1)}} -  \sqrt{1-\Omega_0} 
\over
( 1 - \sqrt{1-\Omega_0} )
}
\right\}^{2/(3 w_*+1)}
\right.
\nonumber\\
\fl
&&
\qquad\qquad
-
\left.
\left\{
{
\sqrt{1-\Omega_0+\Omega_0(1+z)^{(3w_*+1)}} +  \sqrt{1-\Omega_0} 
\over
( 1 + \sqrt{1-\Omega_0} )
}
\right\}^{2/(3 w_*+1)}
\right].
\label{E:dP}
\end{eqnarray}
In this final expression we are always raising quantities to the same power, and the difference between the two terms is just in the placement of $+$ and $-$ signs.  (Note that this expression is guaranteed to be real whatever the value of $\Omega_0$; for $\Omega_0>1$ the two terms are complex conjugates of each other and after taking the pre-factor $\sqrt{1-\Omega_0}$ into account, the overall combination is guaranteed to be real.)

Note that once we have an explicit formula for the (Peebles) angular diameter distance $d_P$, any of the other standard cosmological distances can easily be obtained by multiplying by suitable powers of $(1+z)$~\cite{Peebles, Weinberg, Hogg}, see also~\cite{redshift,controversy}. In particular the luminosity distance is
\begin{eqnarray}
\fl
d_{L_{w_*}}(z) &=&  {c\over2H_0\sqrt{1-\Omega_0} } \left[
\left\{
{
\sqrt{1-\Omega_0+\Omega_0(1+z)^{(3w_*+1)}} -  \sqrt{1-\Omega_0} 
\over
( 1 - \sqrt{1-\Omega_0} )
}
\right\}^{2/(3 w_*+1)}
\right.
\nonumber\\
\fl
&&
\qquad\qquad
\left.
-
\left\{
{
\sqrt{1-\Omega_0+\Omega_0(1+z)^{(3w_*+1)}} +  \sqrt{1-\Omega_0} 
\over
( 1 + \sqrt{1-\Omega_0} )
}
\right\}^{2/(3 w_*+1)}
\right].
\end{eqnarray}
Returning to Peebles' angular diameter distance, the Taylor series expansion in $z$ can be computed as
\begin{eqnarray}
\fl
d_{P_{w_*}}(z) &=& {c\over H_0} \Bigg\{ z  -{2+\Omega_0+3w_* \Omega_0\over 4}\;   z^2 
\nonumber
\\
\fl
&&
+ {4+\Omega_0^2 + w_*(2\Omega_0+6\Omega_0^2) + w_*^2 (-6\Omega_0 + 9 \Omega_0^2 ) \over 8 } \;  z^3 +  \mathcal{O}(z^4) \Bigg\}.\qquad
\end{eqnarray}
Perhaps of more interest is the Taylor series expansion in $\Omega_0$ (since observationally we have good reasons for expecting $\Omega_0\approx 1$). The leading term is easy to calculate
\begin{equation}
d_{P_{w_*}}(z) =  {2c\over H_0(3w_*+1)} \;\left\{1-(1+z)^{-(3w_*+1)/2}\right\}  + \mathcal{O}[\Omega_0-1].
\label{E:easy}
\end{equation}
Extracting the next $\mathcal{O}[\Omega_0-1]$ term is not too difficult, but is somewhat tedious
\begin{eqnarray}
d_{P_{w_*}}(z) &=& {2c\over H_0(3w_*+1)} \;\left\{1-(1+z)^{-(3w_*+1)/2}\right\} 
\nonumber\\
&&
-
{[\Omega_0-1]\; c \over H_0} \Bigg\{ \left[ 
{1-(1+z)^{-(3w_*+1)/2} \over (3w_*+1)}
-
{1-(1+z)^{3(3w_*+1)/2} \over 3(3w_*+1)}  \right]
\nonumber\\
&&
\qquad -
 {1\over 6} \left[ {1 -(1+z)^{-(3w_*+1)/2}\over (3w_*+1)/2 } \right]^3 \Bigg\}
 \nonumber
\\
&&
+\mathcal{O}\left([\Omega_0-1]^2\right).
\label{E:hard}
\end{eqnarray}
In any realistic situation (provided you accept the standard consensus cosmology) the uncertainties in $w$ will completely dwarf any possible effect due to uncertainties in $\Omega_0$, so carrying the expansion to higher order is not warranted.

As usual, $d_{P_{w_*}}(z)$ [or $d_{L_{w_*}}(z)$] can be used to bound $d_{P}(z)$ [or $d_{L}(z)$]. Specifically, let $w$ lie in the range $[w_-,w_+]$ then independent of $\Omega_0\lessgtr 1$:
\begin{itemize}
\item
$d_{P_{w_+}}(z) \leq d_P(z) \leq d_{P_{w_-}}(z); \qquad\qquad  (z > 0)$,
\item
$d_{P_{w_-}}(z) \leq d_P(z) \leq d_{P_{w_+}}(z); \qquad\qquad  (-1< z < 0)$.
\end{itemize}
Here $d_{P_{w_\pm}}(z)$ is given by the rather formidable equation (\ref{E:dP}).

\section{Lookback time}

Finally, consider the ``lookback time'' defined by:
\begin{equation}
\fl
T(z) = \int_a^{a_0} \d t = \int {\d t \over\d a} \d a = \int {a\over \dot a} {\d a\over a} = 
\int {1\over H} {\d(a_0/(1+z))\over a_0/(1+z)} = -\int {1\over H} {\d z/(1+z)^2\over 1/(1+z)}.
\end{equation} 
That is:
\begin{equation}
T(z) = \int_0^z{1\over (1+z) \; H(z)} \; \d z.
\end{equation}
Using the known form of $H_{w_*}(z)$ we define
\begin{equation}
 T_{w_*}(z) \equiv {1\over H_0} \int_0^z{1\over (1+z)^2 \; \sqrt{1+\Omega_0\,((1+z)^{3w_*+1}-1)}}\; \d z,
\end{equation}
and shall use this quantity to place bounds on the actual lookback time $T(z)$.

It is easy to obtain the leading term for $\Omega_0\approx 1$:
\begin{equation}
T_{w_*}(z) = {2\over3\,H_0\,(1+w_*)} \left\{1 - (1+z)^{-(3w_*-1)/2}\right\} + \mathcal{O}[\Omega_0-1].
\end{equation}
The next sub-leading term again is trickier. We eventually obtain
\begin{eqnarray}
\fl
T_{w_*}(z) &=& {2\over3\,H_0\,(1+w_*)} \left\{1 - (1+z)^{-(3w_*-1)/2}\right\}  
\nonumber\\
\fl
&&
-
{[\Omega_0-1]\over H_0} \left[ 
{1-(1+z)^{-3(w_*+1)/2} \over 3(w_*+1)}
-
{1-(1+z)^{-(9w_*+5)/2} \over 9w_*+5}  \right]
\nonumber\\
\fl
&&
+\mathcal{O}\left([\Omega_0-1]^2\right).
\quad
\end{eqnarray}
Again, in any realistic situation (provided you accept the standard consensus cosmology) the uncertainties in $w$ will completely dwarf any possible effect due to uncertainties in $\Omega_0$.
Exact integration and subsequent evaluation of the result for $T_{w_*}(z)$ can only be performed in terms of hypergeometric functions. 
Let us first be a little more careful about the use of the dummy variable in the integration and write
\begin{equation}
\fl
 T_{w_*}(z) = {1\over H_0 \; \sqrt{\Omega_0}} 
 \int_0^z{1\over (1+\tilde z)^{2+1/2(3w_*+1)} \; \sqrt{1-(1-\Omega_0^{-1})\,(1+\tilde z)^{-(3w_*+1)}}}\; \d \tilde z,
\end{equation}
and then, (following the procedure of~\cite{prd,celine-ec}), apply the binomial theorem
\begin{equation}
\fl
 \left[1-(1-\Omega_0^{-1})\,(1+\tilde z)^{-(3w_*+1)}\right]^{-1/2} = 
 \sum_{n=0}^\infty {-1/2\choose n} \;
  (-1)^n \; (1-\Omega_0^{-1})^n\;(1+\tilde z)^{-(3w_*+1)n}\!.\;
\end{equation}
Now this particular binomial series will converge provided~\footnote{Note that for $z<0$ one is actually calculating the ``lookforward time'' --- the time until the universe expands by an additional factor of ${1\over1-|z|}$ in each direction.}
\begin{equation}
\left|  (1-\Omega_0^{-1})\,(1+\tilde z)^{-(3w_*+1)}\right| < 1.
\end{equation}
That is, provided
\begin{equation}
\left|1-\Omega_0^{-1}\right| < (1+\tilde z)^{3w_*+1}.
\end{equation}
More explicitly, the \emph{integral} will make sense provided 
\begin{equation}
\left|{1-\Omega_0\over\Omega_0}\right| < (1+\tilde z)^{3w_*+1};
\qquad  \forall \; \tilde z \in (0,z) \hbox{ or } \tilde z \in (z, 0),
\end{equation}
which is equivalent to
\begin{equation}
\left|{1-\Omega_0\over\Omega_0}\right| <  \min\{ 1, (1+ z)^{3w_*+1} \}.
\end{equation}
\begin{itemize}

\item In all cases, to ensure convergence at redshift zero,  we must certainly have
\begin{equation}
\left|  1-\Omega_0^{-1}\right| < 1, \qquad\hbox{that is}\qquad \Omega_0\in(1/2,\infty).
\end{equation}

\item If $z>0$ and $(3w_*+1) \geqslant0$, ($w_*\geqslant -1/3$), or if $z<0$ and $(3w_*+1) \leqslant0$, ($w_*\leqslant -1/3$):
Then  $(1+z)^{(3w_*+1)}\geq1$, and no additional limitation is imposed.

\item If $z>0$ and $(3w_*+1)<0$, ($w_*<-1/3$), or if  $z<0$ and $(3w_*+1)>0$, ($w_*>-1/3$): 
In this situation  $(1+z)^{(3w_*+1)}<1$, therefore we now obtain an additional limitation on $z$ that is necessary to ensure convergence:
\begin{itemize}
\item 
If $z>0$, then we need
\begin{equation}
z<\left|{\Omega_0-1\over\Omega_0}\right|^{-1/|3w_*+1|} \!\!\! - 1 > 0.
\end{equation}
\item
If $z<0$ then we need
\begin{equation}
z>\left|{1-\Omega_0\over\Omega_0}\right|^{1/(3w_*+1)} \!\!\! - 1 < 0.
\end{equation}
\end{itemize}
\item In view of equation (\ref{E:omega}) these last conditions can also be interpreted as constraints on the $\Omega$ parameter at the redshift one wishes to probe:
\begin{equation}
\left|  1-\Omega_{w_*}(z)^{-1}\right| < 1, \qquad\hbox{that is}
\qquad \Omega_{w_*}(z)\in(1/2,\infty).
\end{equation}

\end{itemize}
Subject to this convergence condition we can integrate term by term, and obtain the convergent series
\begin{equation}
\fl
 T_{w_*}(z) = {1\over H_0 \; \sqrt{\Omega_0}} 
 \sum_{n=0}^\infty {-1/2\choose n} (-1)^n { (1-\Omega_0^{-1})^n\,\left[ 1- (1+z)^{-(3w_*+1)n-3/2(w_*+1)} \right]  \over (3w_*+1)n+3/2(w_*+1)} .
\end{equation}
As a practical matter, for many purposes this series representation may be enough, but we can tidy things up somewhat by first defining
\begin{equation}
S_{w_*}(x) =  \sum_{n=0}^\infty {-1/2\choose n} \; {(-x)^n\over(3w_*+1)n+3/2(w_*+1)},
\end{equation}
in which case
\begin{equation}
\fl
 T_{w_*}(z) = {1\over H_0 \; \sqrt{\Omega_0}}  
 \left\{ S_{w_*}\left(1-\Omega_0^{-1}\right) - (1+z)^{-3/2(w_*+1)} \; S_{w_*}\left( {(1-\Omega_0^{-1})\over(1+z)^{3w_*+1}} \right) \right\}.
\end{equation}
Finally we can recognize that $S_{w_*}(x)$ is itself a particular example of a hypergeometric function,~\footnote{
The classical hypergeometric series is given by 
\[
_2F_1 \left( a, b; c; x\right) = \sum_{n=0}^{\infty} \frac{(a)_n (b)_n}{(c)_n} \; \frac{x^n}{n!},
\]
where $(a)_n = a (a+1) (a+2) É (a+n-1)$ is the rising factorial, or Pochhammer symbol. This series is convergent for $|x| < 1$.} 
and so we can write
\begin{eqnarray}
S_{w_*}(x) 
& =& {1\over3/2(w_*+1)} \; _2F_1\left({1\over2}, {3\over2}\left[{w_*+1 \over 3w_*+1}\right]; {1\over2}\left[{9w_*+5 \over 3w_*+1}\right]; x\right).
\end{eqnarray}
Therefore
\begin{eqnarray}
\fl T_{w_*}(z) &\equiv& {1\over  3/2(w_*+1)\, H_0 \, \sqrt{\Omega_0}}  
\times
 \left\{ 
 {}_2F_1\left({1\over2}, {3\over2}\left[{w_*+1 \over 3w_*+1}\right]; {1\over2}\left[{9w_*+5 \over 3w_*+1}\right]; 1 -\Omega_0^{-1}\right)  
 \right. 
 \nonumber \\
\fl &&\left. - (1+z)^{-3/2(w_*+1)} \;\; 
{}_2F_1\left({1\over2}, {3\over2}\left[{w_*+1 \over 3w_*+1}\right]; {1\over2}\left[{9w_*+5 \over 3w_*+1}\right];  {1-\Omega_0^{-1} \over (1+z)^{3w_*+1}} \right) \right\}.
\label{E:T}
\end{eqnarray}
As usual, $T_{w_*}(z)$  can be used to bound $T(z)$. Specifically, let $w(z)$ lie in the bounded range $w(z) \in [w_-,w_+]$,  then independent of $\Omega_0\lessgtr 1$:
\begin{itemize}

\item
${T_{w_+}}(z) \leq  T(z) \leq {T_{w_-}}(z); \qquad\qquad  (z > 0)$,

\item
${T_{w_-}}(z) \leq T(z) \leq {T_{w_+}}(z); \qquad\qquad  (-1< z < 0)$.

\end{itemize}
Here ${T_{w_\pm}}(z)$ is given by the rather formidable equation (\ref{E:T}), based on the use of hypergeometric functions. 

\section{Special cases and consistency checks}

Useful special cases, and consistency checks we can perform on the formalism, include:
\begin{description}
\item[Dust:] 
For pure dust, $w_+=w_-=0$, we have simple exact results
\begin{equation}
H_{\mathrm{dust}}(z) = H_0 (1+z)\sqrt{1+\Omega_0 z} .
\end{equation}
\begin{equation}
\Omega_{\mathrm{dust}}(z) = \frac{\Omega_0 \left(1+z\right)}{1+\Omega_0 \; z}.
\end{equation}
\begin{equation}
\rho_{\mathrm{dust}}(z) = \rho_0 \left( 1+z \right)^{3}.
\end{equation}
\begin{equation}
d_{P_\mathrm{dust}}(z) =  {2c\over H_0} \left\{ {
(\sqrt{1+\Omega_0z}-1)(\sqrt{1+\Omega_0z}-1+\Omega_0)
\over
(1+z) \Omega_0^2
} \right\}.
\end{equation}
\begin{eqnarray}
T_\mathrm{dust}(z) &=& {1\over H_0(1-\Omega_0)} 
\left\{ 1 - {\sqrt{1+\Omega_0 z}\over 1+z} \right\} 
\\
&+& {\Omega_0\over H_0(1-\Omega_0)^{3/2} } \left\{
\tanh^{-1} {\sqrt{1+\Omega_0 z}\over\sqrt{1-\Omega_0} }
-
\tanh^{-1} {1\over\sqrt{1-\Omega_0}} 
\right\}.
\nonumber
\end{eqnarray}
The only one of these equations for which the $\Omega_0\to1$ limit is even remotely subtle is the lookback time, for which
\begin{eqnarray}
T_{\mathrm{dust},\,\Omega_0=1}(z) &=& {2\over 3 H_0} 
\left\{ 1 - {1\over (1+z)^{3/2}} \right\}.
\end{eqnarray}

\item[Radiation:]
For pure radiation, $w_+=w_- = 1/3$, we have 
\begin{equation}
H_\mathrm{radiation}(z)  = H_0 (1+z)\sqrt{1+\Omega_0 [(1+z)^2-1]} .
\end{equation}
\begin{equation}
\Omega_{\mathrm{radiation}}(z) = \frac{\Omega_0 \left( 1+z\right)^2}{1-\Omega_0+\Omega_0 \left( 1+z\right)^2 }.
\end{equation}
\begin{equation}
\rho_{\mathrm{radiation}}(z) = \rho_0 \left( 1+z \right)^{4}.
\end{equation}
\begin{equation}
d_{P_\mathrm{radiation}}(z) =  {c\over H_0} \left\{ {
\sqrt{1+\Omega_0[(1+z)^2-1]}-1
\over
(1+z) \Omega_0
} \right\}.
\end{equation}
\begin{equation}
T_\mathrm{radiation}(z) = {1\over H_0 (1-\Omega_0)} 
\left[ 1 - {\sqrt{ 1+\Omega_0((1+z)^2-1)}\over 1+z}\right].
\end{equation}
The only one of these equations for which the $\Omega_0\to1$ limit is even remotely subtle is the lookback time, for which
\begin{eqnarray}
T_{\mathrm{radiation},\,\Omega_0=1}(z) &=& {1\over 2 H_0} 
\left\{ 1 - {1\over (1+z)^2} \right\}.
\end{eqnarray}

\item[Cosmological constant:]
For pure cosmological constant $w_+=w_-=-1$. We then obtain (now as equalities rather than inequalities) what would for the NEC have been a set of bounds, such as those presented in~\cite{science, prd, mg8, celine-ec} . (That is, a nonzero cosmological constant is right on the verge of violating the NEC.) 

\end{description}

\bigskip
\noindent
Furthermore, comparing with previous results in the literature:
\begin{itemize}
\item 
For $w_-=-1/3$ one has
\begin{equation}
H(z) \geq H_0 \, (1+z).
\end{equation}
This reproduces the SEC lower bound previously investigated in~\cite{science, prd, mg8, celine-ec}.

\item
For $w_-=-1$ one has
\begin{equation}
H(z) \geq  H_0 \; (1+z)  \sqrt{
(1+z)^{-2}+ [\Omega_0-1]\; \left[ (1+z)^{-2} - 1\right] },
\end{equation}
whence
\begin{equation}
\fl
H(z) \geq  H_0 \;  \sqrt{
1 + [\Omega_0-1]\; \left[ 1 - (1+z)^2\right] } = H_0 \;  \sqrt{
\Omega_0 + [1-\Omega_0]\; (1+z)^2 }.
\end{equation}
This reproduces the NEC lower bound previously investigated in~\cite{science, prd, mg8, celine-ec}.

\item
For $w_+=+1$ we have
\begin{equation}
H(z) \leq  H_0 \; (1+z)  \sqrt{
(1+z)^{4}+ [\Omega_0-1]\; \left[ (1+z)^{4} - 1\right] },
\end{equation}
that is
\begin{equation}
H(z) \leq  H_0 \; (1+z)  \sqrt{
1+ \Omega_0\; \left[ (1+z)^{4} - 1\right] }.
\end{equation}
This reproduces the DEC upper bound previously investigated in~\cite{science, prd, mg8, celine-ec}.

\end{itemize}
%

\section{Conclusions}

In the absence of any detailed understanding of the precise nature of the cosmological equation of state $\rho(p)$ it is useful to examine the question of just how much can be deduced with limited information. 
In this article we have worked in terms of the $w$-parameter $w(z)=p/\rho$, and we have used the idealized case of constant $w_*$ as a ``template'' for comparison purposes with more realistic $w(z)$. Specifically:
\begin{itemize}

\item 
For constant $w_*$ the explicit results for the density $\rho_{w_*}(z)$ and Hubble parameter $H_{w_*}(z)$ are well-known. The explicit result for the $\Omega$ parameter $\Omega_{w_*}(z)$ is less well-known, and the explicit results we have obtained for the angular diameter distance $d_{P_{w_*}}(z)$ and lookback time $T_{w_*}(z)$ appear to be both novel and significant.

\item
More importantly we have seen that these idealized results for constant $w_*$ can be used as the basis for general comparison results that bound the various features of the Hubble flow in the following sense:  If we know that $w(z)\in[w_-,w_+]$ between redshift zero and redshift $z$, then for monotonically evolving generic cosmological quantities $X(z)$ we have derived a number of rigorous bounds of the form
\begin{equation}
X_{w_\pm}(z) \leq X(z) \leq X_{w_\mp}(z),
\end{equation}
where we have explicitly seen that the direction of the inequality depends both on the precise details of the evolution of $X(z)$, and on the redshift range of interest.

\end{itemize}
Finally we point out that all of our bounds have been explicitly calculated for \emph{all} signs of the spatial curvature $k\in[-1,0,+1]$, that is for all $\Omega_0$ (though we have restricted ourselves to the physically very plausible  $\Omega_0>0$). The bounds we have derived are thus both very general and very powerful.

\ack

This research was supported by the Marsden Fund administered by the
Royal Society of New Zealand. 
CC was also supported by a Victoria University of Wellington PhD scholarship.

\section*{References}
\addcontentsline{toc}{section}{References}


\end{document}